\documentstyle[
  aps,prl,epsf
  ]{revtex}
\advance\textheight by 0.22in
\advance\topmargin by -0.10in

\begin{document}
\twocolumn[\hsize\textwidth\columnwidth\hsize\csname
@twocolumnfalse\endcsname
\title{Network of edge states: random Josephson junction array description} 
\author{Leonid P. Pryadko$^*$ and Kar\'en Chaltikian$^\dagger$}
\address{$^*$Department of Physics \& Astronomy, 
  University of California, Los Angeles, CA  90095}
\address{$^\dagger$Department of Physics, Stanford University, Stanford, CA
  94305}
\date{October 26, 1997}
\preprint{cond-mat/9707234}
\maketitle
\begin{abstract}
  We construct a generalization of the Chalker-Coddington network
  model to the case of fractional quantum Hall effect, which describes
  the tunneling between multiple chiral edges.  We derive exact {\em
    local\/} and {\em global\/} duality symmetries of this model, and
  show that its infrared properties are identical to those of
  disordered planar Josephson junction array (JJA) in a weak magnetic
  field, which implies the same universality class.  The zero
  frequency Hall resistance of the system, which was expressed through
  exact correlators of the tunneling fields, is shown to be quantized
  both in the quantum Hall limit and in the limit of perfect Hall
  insulator.
\end{abstract}
] %
\narrowtext %
The QH transitions\cite{Prange-book} provide one of the best testing
grounds for our understanding of quantum critical phenomena.  The main
advantages of this system are the availability of well-controlled
samples, and the variety of field and gate-voltage-driven transitions
within one sample.  Unfortunately, it appears that current theoretical
understanding of the bulk QH effect is lagging behind the variety of
experimentally observed features.  Not only the transitions are
found\cite{
  Exper,Wong-95} %
to have universal scaling exponents and universal critical
conductances\cite{Wong-95,Shahar-95A} as suggested\cite{Kivelson-92}
basing on the bosonic Chern-Simons (CS) model\cite{Girvin-Zhang}, %
but their $I$--$V$ traces at different filling fractions appear to be
algebraically related in a surprisingly wide range of parameters.
Most notable are the reflection symmetry\cite{Shahar-95B} of $I$--$V$
traces, and the quantized Hall insulator\cite{Shahar-95A,Shahar-96A}
(QHI) phenomenon.
\begin{figure}
  \centering\leavevmode %
  \def\epsfsize#1#2{0.52#1}%
  \epsfbox{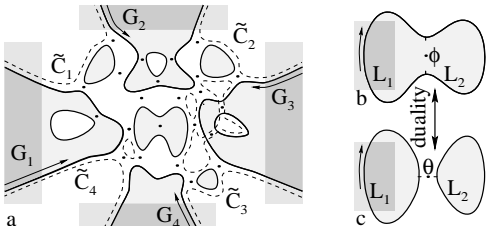}%
    \vskip0.1pc %
  \caption{Left: An illustration of the chiral edge network.  Only the
    tunneling phases in saddle points (solid dots) remain in the
    effective tunneling model.  Four-point measurement can be
    performed via the external gates $G_n$ 
    coupled to long leads.  Dashed lines outline some of the dual
    edges $\tilde{\cal C}_n$. Right: The simplest example of the
    finite-system duality between a QH droplet with a backscattering
    point and two separate droplets connected by a tunneling point.}
  \label{fig:c1}
\end{figure}
Theoretically, despite the attempted\cite{Pryadko-Fradkin} %
rigorous analysis of the bulk CS model, the suggested
universality\cite{Kivelson-92} of the QH transitions has not been
proven beyond the RPA.  It was argued\cite{Shimshoni-96} that both the
charge-vortex duality and the QHI phenomenon (the semicircle
law\cite{Dykhne-Ruzin}) must be present simultaneously to account for
the observed symmetry of $I$--$V$ traces.  Although the derivation of
the semicircle law implies a duality between macroscopic current
density and electric field, so far there was no quantum model to
demonstrate both properties.  Note, however, that the network of
chiral edges, illustrated in Fig.~\ref{fig:c1}a, shows both the correct
scaling behavior (integer regime, Chalker \&
Coddington\cite{Chalker-88}), and the QHI phenomenon (high
temperatures, Auerbach \& Shimshoni\cite{Shimshoni-Auer-96}).
Moreover, the effective action for a single tunneling junction has an
exact {\em dual\/} representation\cite{Schmid-Guinea,Fendley-95}.  An
important question is whether these features hold for the network of
fractional edges at small temperatures.

The goal of this letter is to construct the tunneling action for the
network of chiral edge states.  We derive {\em local\/} and {\em
  global\/} duality symmetries of the partition function, and show
that the infrared properties of the system are identical to those of
the disordered JJA in a weak magnetic field, which implies the same
universality class.  We also express the four-point resistance of this
system through exact correlators of the tunneling fields.

Gapless edge excitations $u\equiv u(x,\tau)$ in the QH regime are
described\cite{Wen} by the imaginary-time action
\begin{equation}
  \label{eq:edge-action}
  S_0=\frac{1}{4\pi}\int{u'(i\dot{u}+ v u' )} dx\,d\tau,
\end{equation}
in our units the edge wave velocity $v\!=\!1$.  The action re\-mains
qua\-drat\-ic\cite{Wen,Hangmo-96} in the presence of Coulomb forces;
the interaction is introduced by the tunneling
\begin{equation}
  \label{eq:tunn-action}
  S_t=\sum_{i<j\le 2 N}\int d\tau \sum_{k=1}^\infty
  \lambda_{ij}^{(k)} \cos(g k \phi_{ij}),
\end{equation}
where $g\!=\!\sqrt\nu$ for the qua\-si\-par\-ti\-cles' tunneling
between the points $x_i$ and $x_j$ through the condensate with the
fill\-ing fraction $\nu\!=\!(2m+1)^{-1}$, or $g\!\rightarrow \!\tilde
g=1/\sqrt{\nu}$ for tunneling of electrons through an insulating
region.  The tunneling amplitudes $\lambda_{ij}^{(k)}$, set by the
details\cite{Jain-88} of the self-consistent potential near saddle
points $\alpha\equiv{ij}$, $1\!<\!\alpha\!\leq\!N$ are treated as
phenomenological parameters.  Although the argument of the cosine in
Eq.~\ref{eq:tunn-action} is usually written as the dif\-fer\-ence
$\phi_{ij}\!=\!u_j\!-\!u_i$, we must remember that tunneling connects
two separate points, and there may be an additional gauge field
contribution\cite{Note-Kane}\nocite{Hangmo-96}.

The non-linear action~(\ref{eq:tunn-action}) depends on values $u_i$
of field $u$ in a discrete set of points $x_i$; its fluctuations in
all other points can be integrated out.  Leaving the arguments
$\phi_\alpha$ of the tunneling terms as the only independent variables
({\em cf}.\ gauge-invariant phases for Josephson junctions), we write
the most general form of the tunneling action
\begin{equation}
  \label{eq:eff-action}
  S[\phi_{\alpha}]=T\sum_{\omega=2\pi n T} {\omega\over4\pi} \bar\phi_\alpha
  K_{\alpha\beta} \phi_\beta+ S_t,
\end{equation}
with Matsubara frequencies $\omega$ and
$\bar{\phi}_\alpha(\omega)\equiv \phi_\alpha(-\omega)$.  Different
tunneling points are linked by the elements of the coupling matrix
$K_{\alpha\beta}(\omega)=-K_{\beta\alpha}(-\omega)$, determined by the
geometry of edge channels; the factor $\omega/4\pi$ was separated for
notational convenience.

Before specifying the form of the coupling matrix $K$, let us discuss
general properties of the action~(\ref{eq:eff-action}):\hfill\\
\noindent ({\bf i}) In the absence of any tunneling the matrix $K$ is uniquely
determined by correlators of the fields $\phi_\alpha$,
\begin{equation}
  \label{eq:zero-lambda-K}
  \bigl(K^{-1}\bigr)_{\alpha\beta}={\textstyle {\omega\over2\pi}}\left\langle
    \phi_\alpha\bar\phi_\beta \right\rangle_{\lambda=0}.
\end{equation}
Therefore, the matrix elements between the tunneling points located at
otherwise disconnected edges vanish.
\noindent({\bf ii})  In the limit of infinite frequencies, the 
coherence between different tunneling points is also lost, while the
matrix $K$ acquires its asymptotic form
\begin{equation}
  \label{eq:asymptotic}
  K_{\alpha\beta}(\omega)
  \stackrel{\omega\rightarrow\infty}{\longrightarrow} 
  \delta_{\alpha\beta}\,{\rm sgn}\,(\omega)
\end{equation}
({\bf iii}) The action~(\ref{eq:eff-action}) describes a complicated
system of non-linear damped anharmonic oscillators.  Since their
mutual coupling is linear, each of these oscillators may be regarded
as a particle
$$
S_1={{T\over4\pi}}\sum_\omega \omega\bigl[
  K_{11}(\omega)|\phi_1|^2+V_1\bar\phi_1-\phi_1\bar V_1\bigr]
+S_t[\phi_1],
$$
in its own periodic potential and external field
$V_1(\omega)=\sum_{\alpha\neq 1} K_{1\alpha}(\omega)\phi_\alpha(\omega)$
due to all other tunneling points, whose dynamics is determined by the
remaining terms $S'[\phi_2,\ldots,\phi_N]$ of the original action.
The partition function can be calculated in two steps,
$$
Z=\!\int\! e^{-S}{\cal D}[\phi]=\int\!{\cal D}\phi_2\ldots {\cal
  D}\phi_N \displaystyle e^{-S'[\phi_2,\ldots,\phi_N]}
Z_1\left[V_1\right],
$$
where $Z_1\left[V_1\right] = \int{\cal D}\phi_1\exp\left(-S_1\right)$.
This one-point partition function can be %
rewritten\cite{Schmid-Guinea}
in terms of the dual variable $\theta_1$ with the action
\begin{equation}
  \label{eq:dual-one-point-action}
  \tilde S_1={T\over4\pi}\sum_\omega{\omega\over
    K_{11}(\omega)}|\theta_1(\omega)+V_1(\omega)|^2+ \tilde
  S_t[\theta_1(\tau)], 
\end{equation}
where the tunneling part $\tilde S_t[\theta]$ has exactly the
form~(\ref{eq:tunn-action}) with modified tunneling coefficients
$\tilde\lambda^{(k)}$ and additional replacement $g\rightarrow\tilde
g\!=\! g^{-1}$.  This duality represents\cite{Fendley-95} a freedom to
describe the same junction in terms of {\em weak\/} tunneling or {\em
  strong\/} backscattering as illustrated in Fig.~\ref{fig:c1}.  If
applicable to larger systems, such duality must severely restrict the
form of the matrix $K$.

We illustrate these properties by calculating the coupling matrix for
systems in Fig.~\ref{fig:c1}b,c.  For the single-edge geometry in
Fig.~\ref{fig:c1}b, the total charge $Q\!=\!\int\!\! \rho(x)\,
dx\!=\!g\,(u_L\!-\!u_0)/2\pi$ is conserved, and it is sufficient to
consider {\em periodic\/} configurations $u_{x+L}=u_x$, where
$L=L_1+L_2$ is the total edge length.  The integration over
fluctuations of the Gaussian field $u$ is performed by evaluating the
action~(\ref{eq:edge-action}) along the classical trajectory
connecting the points $u_j\equiv u(x_j)$ and $u_{j-1}$, $j=1,2$, with
the result
\begin{equation}
  \label{eq:integrated-action}
  S_0[u_j,u_{j-1}]= T\sum_{\omega}
  {\omega\over4\pi}(\bar{u}_{j}-\bar{u}_{j-1})
    {(u_j s_j-u_{j-1})\over{s_j-1}}.
\end{equation}
Here the distance factor $s_j=\exp(\omega L_j)$ depends on the
distance along the edge $L_{j}=x_j-x_{j-1}$.  The only element of the
coupling matrix $K$ is
\begin{equation}
  \label{eq:K-matr-c1a}
  K^{(\ref{fig:c1}b)}_{11}={s_1 s_2-1\over(s_1-1)(s_2-1)}\equiv
  {e^{\omega (L_1+L_2)}\!-\!1\over(e^{\omega L_1}\!-\!1) (e^{\omega
      L_2}\!-\!1)}.
\end{equation}
Clearly, the property ({\bf ii}) is satisfied: at high enough
frequencies, $|\omega| L_{1,2}\gg 1$, the
expression~(\ref{eq:K-matr-c1a}) reduces to ${\rm sign}(\omega)$, up
to exponentially small corrections.

The two-edge geometry shown in Fig.~\ref{fig:c1}c is homomorphic to
that of QH liquid on the surface of a cylinder investigated by
Wen\cite{Wen}, who emphasized that mere collection of independent
edges does not give its complete description\cite{Note-selection}.
This is related to the fact that tunneling destroys the charge
conservation at individual edges, and the periodic boundary conditions
are no longer valid.  It turns out that the free boundary conditions
($u_L$ independent of $u_0$) combined with the ``boundary terms''
\begin{equation}
  S_{\rm end}= T/(8\pi)\textstyle\sum_\omega {\omega} {(\bar u_L+\bar
    u_0)}(u_L-u_0),
  \label{eq:end-correction}
\end{equation}
which must be added to the action~(\ref{eq:edge-action}) for every
separate edge, restore the correct Hilbert space.  Particularly, for
the system in Fig.~\ref{fig:c1}c, after integrating out all Gaussian
modes, this prescription leads to the effective action
\begin{equation}
  \tilde S=T\sum_\omega {\omega|\theta_\omega|^2\over4\pi}
  {(e^{\omega L_1}\!-\!1)(e^{\omega L_2}\!-\!1) %
    \over e^{\omega (L_1+L_2)}\!-\!1}+S_t[\theta],
  \label{eq:act-c1b}
\end{equation}
which satisfies the finite-size duality requirement ({\bf iii}).

Although the same prescription works for larger networks, in systems
with more than four tunneling points direct calculation of $\hat K$
becomes too bulky.  Instead we use Eq.~(\ref{eq:zero-lambda-K}), which
implies that the element of the inverse coupling matrix
$K^{-1}_{\alpha\beta}$ is independent of the presence of the tunneling
points $\gamma\neq\alpha,\beta$.  The diagonal elements of matrix
$\hat K^{-1}$ can be always deduced from Eqns.~(\ref{eq:K-matr-c1a})
or~(\ref{eq:act-c1b}) by substituting the appropriate lengths for
$L_1$ and $L_2$, while the off-diagonal ones can be derived by
constructing effective actions with only {\em two\/} tunneling points.
\begin{figure}
  \centering \leavevmode %
  \def\epsfsize#1#2{0.52#1}%
  \epsfbox{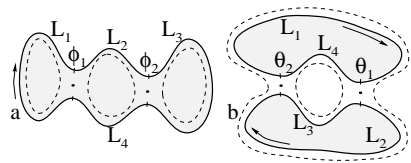}\vskip0.01pc %
  \caption{Non-trivial geometries with two tunneling points.}
  \label{fig:c2}
\end{figure}
The coupling matrices for two-tunneling-point geometries shown in
Fig.~\ref{fig:c2} are given by the expressions
\begin{eqnarray}
  \label{eq:K-matr-c2a}
  \hat K^{\rm (\protect\ref{fig:c2}a)}&=&\displaystyle\left|
  \begin{array}[c]{cc}
    {s_1 s_2 s_4-1\over(s_1-1)(s_2 s_4-1)}&
    {s_2\over s_2 s_4-1}\\
    {s_4\over s_2 s_4-1}& 
    {s_2 s_3 s_4-1\over(s_3-1)(s_2 s_4-1)}
  \end{array}\right|,\\
  \label{eq:K-matr-c2b}
  \hat K^{\rm (\protect\ref{fig:c2}b)}&=&\displaystyle\left|
    \begin{array}[c]{cc}
      \textstyle {s_1 s_2 s_3 s_4-1\over (s_1 s_2-1)(s_3 s_4-1)}&
      { s_4\over  1-s_3 s_4}+{s_2\over 1-s_1 s_2}\\
      {s_3\over 1-s_3 s_4}+{s_1\over 1-s_1 s_2}&
      {s_1 s_2 s_3 s_4-1\over (s_1 s_2-1)(s_3 s_4-1)}
    \end{array}\right|,
\end{eqnarray}
where $s_i=\exp (\omega L_i)$.  All other geometries with two
tunneling points can be generated from these by replacing one or both
tunneling junctions by their dual(s) as illustrated in
Fig.~\ref{fig:c1}b,c.  Our calculation shows that the coupling matrices
for thus obtained configurations are immediately related to
Eqns~(\ref{eq:K-matr-c2a}), (\ref{eq:K-matr-c2b}) as prescribed by
Eq.~(\ref{eq:dual-one-point-action}): the modified coupling matrix is
given by the coefficients of the Legendre-transformed bilinear form
\begin{equation}
  \label{eq:legendre-defined}
  \tilde S_0[\theta_1,\phi_2]\!=\!{T\over4\pi}\!\!\sum_\omega \omega{
    \bigl(\bar\phi \hat K\phi\! -\!
      \bar\phi_1\theta_1\!+\!\bar\theta_1\phi_1\bigr)}
  \bigr|_{K_{1\alpha}\phi_\alpha=\theta_1}  .
\end{equation}
This implies that under the {\em global\/} duality transformation,
which replaces {\em all\/} tunneling junctions by their duals, the
coupling matrix is inverted, $\tilde K=K^{-1}$.  Combining this
statement with Eq.~(\ref{eq:zero-lambda-K}), we find that the original
matrix $K$ is given by the correlators of the {\em dual\/} fields
\begin{equation}
  \label{eq:zero-tilde-lambda-K}
  K_{\alpha\beta}= {\omega\over2\pi}\left\langle
    \theta_\alpha\bar\theta_\beta\right\rangle_{\tilde\lambda=0}
\end{equation}
in the limit of perfect transmission (zero backscattering) in every
tunneling point.

Although the dual form of the model is the result of a non-local
change of variables in the path integral, the {\em weak\/}-coupling
limits of the original model and its dual version can be interpreted
as describing the same system at mutually dual filling fractions, with
the total areas of the condensate and the depleted regions
interchanged.  This interpretation is correct as long as the
probability distribution of the bare parameters of the original model,
and the model at the dual filling fraction expressed in dual variables
are similar, which is expected if the distribution of the disorder
potential is
symmetric.  
Under such duality ``in average'' $g\!\rightarrow\!g^{-1}$, and
therefore at $g\neq1$ this transformation is {\em not\/} an exact
symmetry of the problem.  This is obvious in the limit of large
frequencies, where the off-diagonal elements of the coupling matrix
$\hat K$ vanish, and the scaling dimensions of the tunneling
amplitudes $\lambda_{ij}^{(k)}$ coincide with their values
$\varepsilon_k\!=\!1\!-\!k^2 g^2$ for isolated tunneling junctions.

At small frequencies, however, the interference becomes important, and
the scaling behavior crosses over to mesoscopic regime where the
quantization of the drift orbits is relevant.  It turns out that the
quadratic part of the action in this limit can be written as a simple
sum
\begin{displaymath}
  \lim_{\omega\rightarrow0} \omega K_{\alpha\beta}
  \bar\phi_\alpha\phi_\beta=\sum_{\tilde{\cal C}}{1\over L_{\tilde{\cal
        C}}} \left|\Phi_{\tilde{\cal C}}\right|^2,
\end{displaymath}
where 
\begin{math}
 \Phi_{\tilde{\cal C}}\equiv %
 \sum_{\alpha\subset\tilde{\cal C}}\phi_{\alpha}
\end{math}
is the directed sum of the tunneling phases $\phi_\alpha$ along the
dual edge ${\tilde{\cal C}}$ of length $L_{\tilde{\cal C}}$ created by
the global duality transformation.  The presence of such terms in the
action imply that variables $\Phi_{\tilde{\cal C}}$ can have only
finite r.m.s.~deviation, which ensures that the averages
\begin{math}
  \langle\dot\Phi_{\tilde{\cal C}}\rangle=\sum_{
    \alpha\subset{\tilde{\cal C}}}\langle\dot\phi_{\alpha}\rangle=0
\end{math}
vanish.  These constrains can be resolved by writing the phases
$\phi_\alpha$ as the gradients $\phi_\alpha=g\Delta_\alpha{\cal
  V}+{\cal A}_\alpha$ of the local potential ${\cal V}\equiv {\cal
  V}_{{\cal C}}$, associated with every independent edge ${{\cal C}}$,
up to some time-independent phases ${\cal A}_\alpha$.  In this
representation the model becomes identical to a disordered JJA in
external magnetic field.  Similarly, in the dual representation of the
model the sum $\dot\Theta_{\cal C}=\sum_{\alpha\subset{\cal C}}
\dot\theta_\alpha$ can be associated with the total current of
composite bosons entering the enclosed area; this current also
vanishes at zero frequency.

The performed analysis of the infrared limit reveals a deep analogy
between the CB in disordered QH systems and bosons near
disorder-driven superconductor--insulator (SC--I) transition.  After
the system is separated into weakly coupled phase-coherent areas, the
terms remaining in small-fre\-quen\-cy expansion of the action depend
on lengths $L_\alpha$ in a highly non-universal fashion.  This is used
to absorb the Luttinger liquid coupling constant $g$, which renders
the tunneling model independent of the filling fraction $\nu$, leading
to the conclusion that all QH transitions must be in the same
universality class, associated with disordered SC--I transition.

This universality follows only from the mapping between the partition
functions, and it does not imply the identical transport properties.
The chiral nature of the current-carrying states in the edge network
model is responsible for its large Hall resistance. This can be
readily seen in the integer regime using the
B\"uttiker-Landauer\cite{Buttiker-86} formula relating the four-point
resistance tensor with the matrix $T_{nn'}\!=\!|t_{nn'}|^2$,
$n,n'\!=\!1,\ldots,4$ of transmission probabilities between the
incoming and outgoing channels as illustrated in Fig.~\ref{fig:c1}a.
Although for an arbitrary unitary scattering matrix $t_{nn'}$ the Hall
resistance $R_{\rm H}=(R_{xy}-R_{yx})/2$ is not necessarily quantized,
in the {\em classical\/} regime, which is characterized by the absence
of the quantum interference between different paths contributing to
$t_{nn'}$, $R_{\rm H}=\pm1$, in agreement with
Ref.\CITE{Shimshoni-Auer-96}.

Within the proposed model, the infinite ideal leads ${\cal C}_n$ can
be coupled with the potential gates $G_n$ using the action
$S_{G_n}\!=\!2\pi\,i\,g^{-1}\!\int\!  d\tau\,\Theta_n W_n$, where the
phase $\Theta_n\!=\!\sum_{\alpha\subset {\cal C}_n}\theta_\alpha$ is
proportional to the total charge transferred to the lead ${\cal C}_n$,
and $W_n$ is the potential of the corresponding gate $G_n$.  In the
linear responce regime the current to the lead $n$ is given by the
expression
\begin{equation}
  \label{eq:current-one}
  {\cal I}_n(\omega) 
  ={\omega g^{-2}}
  \langle\Theta_n\bar\Theta_{n'}\rangle W_{n'}(\omega),
\end{equation}
while the equilibrium potential differences between the opposite leads
can be evaluated by decomposing them into individual chiral channels,
{\em e.g.\/}
\begin{equation}
  \label{eq:voltage-formula}
  2 g ({V}_1\! -\!{V}_3)+ g^{-1}({\cal I}_2\!-\!{\cal I}_4)=
  \tilde{\cal I}_1\!-\!\tilde{\cal I}_2\!-\!\tilde{\cal
  I}_3\!+\!\tilde{\cal I}_4. 
\end{equation}
The currents
\begin{equation}
  \label{eq:current-dual-one}
  \tilde{\cal I}_n(\omega)
  ={\omega}
  \langle\Phi_n\bar\Theta_{n'}\rangle W_{n'}(\omega),
\end{equation}
are defined at the external edges $\tilde{\cal C}_n$ (dashed lines in
Fig.~\ref{fig:c1}a) in the {\em dual\/} representation of the model.
The r.h.s.\ of Eq.~(\ref{eq:current-dual-one}) can be related to the
potential difference between the phase-coherent regions of leads 1 and
3 in the effective JJA model, which leads to the formula
\begin{equation}
  \label{eq:resistance-decomposed}
  \hat R\!=\!-g^{-2}\hat\Sigma\!+\!\hat R_{\rm b},
\end{equation}
for the resistance tensor as the sum of the quantized Hall resistance
and the part $\hat R_{\rm b}$ associated with the transport of bosons
in the JJA.  An analogous decomposition $\hat\rho=\rho_{\rm
  quant}+\hat\rho_{\rm b}$ was previously derived\cite{Kivelson-92}
for resistivities in the RPA approximation; we believe this expression
is exact here because the charges are essentially localized by
disorder and the fluctuations of the Chern-Simons field are
suppressed.  The usual relationship $\hat\rho_{\rm
  b}\rightarrow\hat\sigma_{\rm b}^{\rm t}$ implied by the
particle-vortex duality\cite{Fisher-90,Kivelson-92,Pryadko-Fradkin} in
bulk models, is valid in the considered model only after averaging
over the disorder.  

Eliminating auxiliary potentials $W_n$ from
Eqns.~(\ref{eq:current-one})--(\ref{eq:current-dual-one}) and an
expression similar to Eq.~(\ref{eq:voltage-formula}) for leads 2 and
4, we finally obtain
\begin{equation}
  \label{eq:R-matrix-fract}
  \hat R_{\rm b}\!=\!g^{2}
  (\hat1\!+\!\hat\Sigma) \bigl[{\textstyle {1\over2}}
    \hat D\,\langle\Phi\bar\Theta\rangle\,
  \langle\Theta\bar\Theta\rangle^{-1}\, \hat D^{\rm t}\bigr],
\end{equation}
where the matrices 
\begin{equation}
  \label{eq:matrices-defined}\textstyle
  \hat\Sigma = \biggl|
    \begin{array}[c]{rr} 0&1\\-1&0  \end{array}\biggr|,\quad
  \hat D = \biggl|
  \begin{array}[c]{rrrr} 1&0&-1&0\\0&-1&0&1  \end{array}\biggr|,
\end{equation}
represent the configuration of currents and potential differences in
the four-lead measurement.  Note that after the rescaling $g\,\Phi
\rightarrow\Phi$, $\Theta/g\rightarrow\Theta$ used to construct the
effective JJA model, the expression~(\ref{eq:R-matrix-fract}) loses
its explicit dependence on the filling fraction $\nu=g^2$.  

The correlator $\langle\Phi\bar\Theta\rangle$ in
Eq.~(\ref{eq:current-dual-one}) is the linear combination of the
expressions
\begin{displaymath}
  \langle\phi_\alpha\bar\theta_\beta\rangle 
  \!=\!{\textstyle{2\pi\over\omega}}\delta_{\alpha\beta}
  \!-\!  \langle\phi_\alpha\bar\phi_\gamma\rangle K_{\gamma\beta}
  \!=\!-{\textstyle{2\pi\over\omega}}\delta_{\alpha\beta}
  \!+\!
  K_{\alpha\gamma}^{-1}\langle\theta_\gamma\bar\theta_\beta\rangle.
\end{displaymath}
According to Eqns.~(\ref{eq:zero-lambda-K}) and
(\ref{eq:zero-tilde-lambda-K}), these vanish identically if the
tunneling is absent in either the original ($\lambda_\alpha\!=\!0$) or
the dual ($\tilde\lambda_\alpha\!=\!0$) representations, which leaves
only the quantized part of the Hall resistance.

To conclude, we constructed the effective tunneling model which
generalizes the Chalker-Coddington model\cite{Chalker-88} to the
fractional QH effect.  The partition function of this model can be
approximately mapped to that of the disordered JJA, which implies the
universality of the quantum Hall transitions.  The model always has an
exact dual representation, but it maps to the same system at different
filling fraction only in the limit of small frequences and after
averaging over disorder.  The associated relationships between the
filling fractions and the transport coefficients are identical to
those obtained in the bulk CS model.  Although in the limit of large
temperatures the constructed model becomes equivalent to classical
resistor network, which demonstrates\cite{Shimshoni-Auer-96} the QHI
phenomenon, in general this behavior may not be present.  Finally,
since the constructed model involves no additional approximation
compared with the full chiral Luttinger network model, it can be used
to simulate tunneling experiments in the fractional QH regime.

It is our pleasure to acknowledge beneficial discussions with
S.~Chakravarty, E.~Fradkin, I.~Gruzberg, C.~Kane, S.~Ki\-vel\-son,
D.-H.~Lee, C.~Marcus, E.~Shimshoni and S.-C.~Zhang.


\begin{thebibliography}{88}
\bibitem{Note-Kane} This contribution was not taken into account in
  Ref.\protect\CITE{Hangmo-96}.  
\end{thebibliography}
\end{document}